\documentstyle[12pt]{article}

\newcommand{\ber}{\begin{eqnarray}}
\newcommand{\eer}{\end{eqnarray}}
\newcommand{\bea}{\begin{equation}}
\newcommand{\eea}{\end{equation}}
\begin{document}
\title{\bf Two Time Correlation For A Noise Driven Double-Well Oscillator In The Suzuki Regime}
\author {\bf Suvadeep Bose${}^1$ and  Saumyadip Samui${}^2$ \\
        Department of Physics, Jadavpur University\\
        Jadavpur, Calcutta - 700032, India.\\
	e - mail: ${}^1$suvadeepbose@yahoo.com, 
                  ${}^2$ssamui2000@yahoo.com }
\date{}
\maketitle
\begin{abstract}
We study the two time correlation for the noise driven dynamics of the double-well oscillator in the Suzuki regime. It is seen that for very small noise strength the correlation function shows a lack of translational invariance for very long times characterised by the Suzuki scaling variable. We see that in this strongly out-of-equilibrium situation, the conventional mode-coupling approximation is not a convenient tool. 
\end{abstract}
\vspace{3.0cm}
\indent PACS:\hspace{1.0cm}    05.40.-a , 05.20.-y
\vspace{0.5cm}

\newpage

Non-equilibrium, time dependent correlation functions generally show a breakdown of time translation invariance${}^1$. This is very strongly seen in various glassy${}^{2,3}$ systems where a phenomenon called `aging' occurs. In this phenomenon it is noticed that the a.c. susceptibility depends both on frequency and time. The response of the system to a thermal perturbation thus depends on thermal history. If the `aging' effect were absent, then the a.c. susceptibility would depend on $\omega$ alone which implies that the two time correlation function depends only on the time difference. The fact that the a.c. susceptibility depends on the time elapsed is a reflection of the two time correlation function depending on the time difference as well as the actual time elapsed. In this work we consider another problem which has received a lot of attention - the approach to equilibrium in a double well potential. The time scale for this equilibration is exponentially large and consequently we expect the breaking of time translation invariance in the two time correlation function for a long time. In the double well oscillator problem, Suzuki scaling${}^{4,5}$ is one of the most important observations and in this work, we shall concentrate on the two time correlation in the Suzuki regime. What we shall see is that in this very simple system, the two time correlation depends on the initial time as well as the time difference.  
 
The system studied by Suzuki was the stochastic differential equation

\begin{equation}
\dot{x}=\gamma x-g x^{3}+\eta(t)
\end{equation}

where $\eta(t)$ is a Gaussian white noise whose correlation is given by

\begin{equation}
<\eta(t)\eta(s)>=2\epsilon\delta(t-s)
\end{equation}

\indent In the above equation $\gamma$ and $g$ are both positive, so that the correlation function runs away for $g=0$. This means that the problem is not amenable to pertubative treatment. What Suzuki had explored was the time dependance of $ < x^{2}(t)>$ and the manner in which it eventually reached its equilibrium value $<x^{2}>_{st}$. What was found was that

\bea
<x^{2}(t)>=<x^{2}>_{st}\lbrace 1-\frac{2}{\sqrt{2\pi}}\int_0^{\infty}\frac{e^{-\xi^2/2}}{1+\xi^2\tau_{0}} d\xi\rbrace
\eea

where $\tau_{0}=\frac{g \epsilon}{\gamma^2} e^{2\gamma t}$. The combination $\tau_{0}$ which plays the role of a scaled time determines the dynamics in the Suzuki regime and we see that large time means large $\tau_{0}$, and this can be achieved only if $ t>>0 \hspace{0.2cm} (\ln{\frac{1}{\epsilon}})$.

 One of the ways of obtaining this result was to sum an infinite series of terms in perturbation theory. In this work we would like to look at the two-time correlation function 
$C(t_1,t_2)=<x(t_1)x(t_2)>$. Our reason for looking at this correlation function is to point out the possible inequivalence of different calculational tools when an out-of-equilibrium situation needs to be handled. One of the well-known schemes for calculating correlation function is the mode-coupling scheme. In this scheme, one needs a dressed Green's function $G(t_1,t_2)$ and a self energy $\Sigma(t_1,t_2)$ in addition to the correlation function $C(t_1,t_2)$. In this scheme 

\bea
C(t_1,t_2)=2\epsilon\int G(t_1,t_3)G(t_3,t_2)dt_3+6g^2\int G(t_1,t_3)\lbrace C(t_3,t_2)\rbrace ^3 G(t_3,,t_2)dt_3
\eea

\bea
G(t_1,t_2)=G(t_1,t_2)+\int \int G_0(t_1,t_3)\Sigma(t_3,t_4)G(t_4,t_2)dt_3dt_4
\eea

\bea
\Sigma(t_3,t_4)=g^2G(t_3,t_4)[C(t_3,t_4)]^2
\eea

It is obvious that for $g\ll1$, the first correction to the correlation function is $O(g^2)$. If we set $t_1=t_2$ in the correlation function $C(t_1,t_2)$, we should be getting $<x^2(t)>$ for which the first correlation according to Suzuki's formula (Eq.(3)) is $O(g)$. The mode-coupling approximation and the Suzuki type of approximation are consequently very different. This illustrates the difficulty that one encounters in calculating two-time correlation in out of equilibrium situations. Our claim is that any approximation for $C(t_1,t_2)$ must reproduce the equal time function correctly. This is a fairly trivial requirement in near-equilibrium dynamics, but obviously not so easily implemented in the far from equilibrium situations. In Sec II, we explore the spherical limit, where this requirement can be implemented in a straight forward manner.

\newpage

{\bf II :  Spherical Limit}

\vspace{1.0cm}

In this section we generalise $x(t)$ to the $N$-component vector $\bar x(t)$ with components $x_i(t)$, $i$ ranging from 1 to $N$ ${}^{6,7}$. The equation of motion for the $i$-th component is written as 

\bea
\dot x_i =\gamma x_i-\frac{g}{N}\hspace{0.08cm} [\Sigma_{j=1}^{N}x_j^2]\hspace{0.08cm} x_i+\eta(t)
\eea

Multiplying by $x(t+T)$ and taking appropriate averages and making the right symmetrization

\ber
\frac{d}{dt}<x(t+T)x(t)> & = & 2\gamma 
C(t,T)-g\lbrace <x^2(t)>+<x^2(t+T)>\rbrace C(t,T)\nonumber \\
& - & \frac{g}{N}[<x^3(t)x(t+T)>+<x^3(t+T)x(t)>\nonumber\\ 
& - & N\lbrace <x^2(t)>+<x^2(t+T)>\rbrace C(t,T)] \nonumber\\
& + & <\eta(t)x(t+T)>+<\eta(t+T)x(t)>
\eer

In the spherical limit i.e. $N \to \infty$, we note that the quantity \hspace{3.0cm} $<x^3(t)x(t+T)>-N<x^2(t)>C(t,T)$ is $O(1)$ and hence the term involving this quantity is $O(N^{-1})$. The two time correlation function in the spherical limit is

\ber
\frac{d}{dt} C(t,T) & = & 2\gamma C(t,T)-g\lbrace <x^2(t)>+<x^2(t+T)>\rbrace C(t,T)\nonumber\\
& + & <\eta (t)x(t+T)>+<\eta (t+T)x(t)>
\eer   

In the limit $T \to 0$, one correctly recover the spherical limit answers for $S=<x^2(t)>$, which reads 

\bea
\frac{d}{dt}S(t)=2\gamma S(t)-2gS^2(t)+2<\eta(t)x(t)>
\eea
in accodance with reference [6].

Investigating the equation of motion for $<x(t+T)\eta(t)>$ in the spherical limit, we find for $\epsilon \to 0$, the equations

\bea
\frac{d}{dt} C(t,T)=2\gamma C(t,T)-g(S(t)+S(t+T))C(t,T)+2\epsilon
\eea

\bea
\frac{dS}{dt}=2\gamma S-2gS^2+2\epsilon
\eea

Keeping in mind that $\epsilon \to 0$,we can write the solution of Eq.(12) as

\bea
\frac{S}{S_{eq}}=\frac{\tau}{1+\tau}
\eea

where $S_{eq}$ is the equilibrium value of $S$ obtained from the root of 

\bea
2\gamma S-2gS^2+2\epsilon=0
\eea

Clearly, $S_{eq}=\frac{\gamma}{g}+O(\epsilon)$. The time variable $\tau$ is $\epsilon e^{2gS_{eq}t}$, where the boundary condition matching is done in accordance with the Suzuki picture.

Inserting the above $S(t)$ in Eq.(11) and integrating for $\epsilon \to 0$, we find 

\bea
C = \frac{\tau}{(1+\tau)^{1/2}(1+\tau e^{2\gamma T})^{1/2}}
\eea 

where we have restricted ourselves to time-difference $T$ for which \hspace{2.0cm} $\epsilon e^{2\gamma T}\ll1$. If this condition is relaxed, we need to work to greater accuracy in $\epsilon$ and the result is

\bea
C=(\frac{\tau}{1+\tau})^{\frac{\gamma}{2gS_{eq}}}(\frac{\tau e^{2gS_{eq}T}}{1+\tau e^{2gS_{eq}T}})^{\frac{\gamma}{2gS_{eq}}}[1+\frac{1}{e^{2gS_{eq}T}}]^\frac{\gamma}{2gS_{eq}}-(\frac{1}{2})^{\frac{\gamma}{2gS_{eq}}}
\eea

where $S_{eq}$ is now the full expression for the equilibrium correlation function.

We can see from Eq.(13) that as $\tau \to \infty$, i.e. the waiting time becomes very large the correlation function depends on the time difference $T$ alone which we call the stationary state. It is clear that the correlation function splits into two parts - the stationary part and a correction which is explicitly dependant on the waiting time. Thus the spherical limit provides us with a two point correlation function which has a smooth passage to the limit of equal time correlation function. 
\newpage

{\bf III : Summing the Perturbation series in the Suzuki Regime}

\vspace{1.0cm}

In this section, we explore how to generalize Suzuki's technique to the case of time dependent correlation function. To do this we first re-examine Suzuki's result in the light of peturbation theory. Consequently, we expand 

\bea
x=x_0(t)+gx_1(t)+g^2x_2(t)+...
\eea

and insert this in Eq.(1). Equating equal powers of $g$ on either side

\bea
x_0(t)=e^{\gamma t}\int_0^t e^{-\gamma t^{'}}\eta(t^{'})dt^{'}
\eea 

where the initial value of $x_0$ is taken to be zero without loss of generality. Preceeding further we have 

\bea
x_1(t)=-e^{\gamma t}\int_0^t x_0^3(t^{'})e^{-\gamma t^{'}}dt^{'}
\eea

and 

\ber
x_2(t)&=&-3 e^{\gamma t}\int_{0}^{t}dt{'}x_0^{2}(t{'})x_1(t{'})e^{-\gamma t{'}}dt{'}\nonumber\\
      &=&3 e^{\gamma t}\int_{0}^{t}dt{'}x_0^2(t{'})\int_{0}^{t{'}}x_0^{3}(t{''})e^{-\gamma t{''}}dt{''}
\eer

If we now look at the perturbation series for $S(t)=<x^2(t)>$, we note that the Suzuki result is reproduced if every term in the diagrammatics is kept and evaluated to the leading order. Consequently, the two time correlation function $C(t_1,t_2)$ will give the correct $t_1=t_2$ limit only if all the terms in the diagrammatic perturbation theory are kept and evaluated to the leading order. After straightforward but lengthy algebra, we find that to $O(g^2)$

\ber
C(t,t+T)&=&\frac{\epsilon}{\gamma}e^{2\gamma t+\gamma T}\lbrace 1-\frac{3g\epsilon}{\gamma^2}e^{2\gamma t+\gamma T}\cosh(\gamma T)
+\frac{15g^2\epsilon^2}{\gamma^4}e^{4\gamma t+2\gamma T}\cosh^2(\gamma T)+...\rbrace \nonumber\\
& & \lbrace 1+\frac{15g^2\epsilon^2}{2\gamma^4}e^{4\gamma t+2\gamma T}\sinh^2(\gamma T)+...\rbrace 
\eer

The above expression correctly reproduces the Suzuki answer for S(t) when $T = 0$. If we work in the limit of sufficiently small $\epsilon$ and $g$ ( deep well and very little noise ), then for not too large $\gamma T$, we can drop the correction involving $\sinh^2(\gamma T)$ and perform the sum to obtain

\bea
C(t,t+T)=\frac{\gamma}{g}\frac{1}{\cosh(\gamma T)}[1-\frac{2}{\sqrt{2\pi}}\int_0^{\infty} \frac{\xi^2e^{\xi^2/2}}{1+\xi^2 {\cal T}}d\xi]
\eea

where ${\cal T}=\frac{g\epsilon}{\gamma^2}e^{2t+T}\cosh(\gamma T)$.

As in the spherical limit of the previous section, we find that $C\propto e^{-\gamma T}$ after the system has waited for a sufficently long time. The above expression for the two time correlation function is valid in the Suzuki regime i.e. where the system is essentially sliding down the slope of the double well in its attempt to reach equilibrium. It appears that the mode coupling approximation is not an appropriate approximation in this range.

It is for sufficiently long times, when the effect of the noise is again very important in setting up the equilibrium distribution that the mode coupling approximation is going to be useful. This is a situation where $S(t)$ is very nearly constant and now the dynamics of the two-time correlation is determined  by the time difference.

In conclusion, we have seen that for a clearly out-of-eqilibrium situation characterised by the Suzuki regime in a double-well oscillator either the spherical limit or the leading order summation of all diagrams provide a significantly better description of the two-time correlation than the mode-coupling approximation.

\end{document}